\documentclass[twocolumn,preprintnumbers,amsmath,amssymb]{revtex4}

\usepackage{amsmath}    
\usepackage{graphicx}   
\usepackage{color}

\begin{document}

\title{The $\delta$-quantum machine, the $k$-model, and the non-ordinary spatiality of quantum entities}
\author{Massimiliano Sassoli de Bianchi}
\affiliation{Laboratorio di Autoricerca di Base, 6914 Carona, Switzerland}\date{\today}
\email{autoricerca@gmail.com}   

\begin{abstract}

The purpose of this article is threefold. Firstly, it aims to present, in an educational and non-technical fashion, the main ideas at the basis of Aerts' \emph{creation-discovery view} and \emph{hidden measurement approach}: a fundamental explanatory framework whose importance, in this author's view, has been seriously underappreciated by the physics community, despite its success in clarifying many conceptual challenges of quantum physics. Secondly, it aims to introduce a new quantum machine -- that we call the \emph{$\delta$-quantum machine} -- which is able to reproduce the transmission and reflection probabilities of a one-dimensional quantum scattering process by a Dirac delta-function potential. The machine is used not only to demonstrate the pertinence of the above mentioned explanatory framework, in the general description of physical systems, but also to illustrate (in the spirit of Aerts' $\epsilon$-model) the origin of classical and quantum structures, by revealing the existence of processes which are neither classical nor quantum, but irreducibly intermediate. We do this by explicitly introducing what we call the \emph{$k$-model} and by proving that its processes cannot be modelized by a classical or quantum scattering system. The third purpose of this work is to exploit the powerful metaphor provided by our quantum machine, to investigate the intimate relation between the concept of \emph{potentiality} and the notion of \emph{non-spatiality}, that we characterize in precise terms, introducing for this the new concept of \emph{process-actuality}.


\end{abstract}

\maketitle

\section{Introduction}
\label{intro}

Since its origin, it has been known that quantum mechanics (QM) has significant structural differences compared to classical mechanics (CM). For instance, quantum observables, contrary to classical ones, do not necessarily commute and, therefore, are not necessarily (experimentally) compatible. In other terms, the algebra of quantum observables, contrary to the algebra of classical ones, is non-commutative~\cite{Segal, Emch}. 

Also, in CM  propositions (i.e., the statements about the properties of a physical system) are either true or false, and can coherently be combined by means of the disjunction (OR) and conjunction (AND) logical operations, giving rise to a propositional Boolean algebra for which the distributivity property between the OR and AND operations holds. On the other hand, not only quantum propositions are not in general \emph{a priori} either true or false, but also have the tendency to violate the distributivity law; hence, they do not form a Boolean algebra~\cite{Birkhoff, Jauch, Piron}. 

Another important structural difference between QM and CM is the fact that the probability model describing a classical system is Kolmogorovian (i.e., it obeys Kolmogorov's axioms of classical probability theory) whereas the one describing a quantum one is not~\cite{Foulis, Randall, Gudder, Accardi, Pitovski}.

These deep structural differences between QM and CM have certainly contributed to the consolidation of the preconception that QM, contrary to CM, cannot be understood, as exemplified in a Feynman's celebrated quote~\cite{Feynman}: ``There was a time when the newspapers said that only twelve men understood the theory of relativity. I do not believe that there ever was such a time. There might have been a time when only one man did, because he was the only guy who caught on, before he wrote his paper. But after people read the paper a lot of people understood the theory of relativity in some way or other, certainly more than twelve. On the other hand, I think I can safely say that nobody understands quantum mechanics.'' 

One of the purposes of this paper is to show that this preconception is unfounded, as we dispose today of a very clear explanation of the origin of quantum structures. Such an explanation is contained in Aerts' \emph{creation-discovery view}~\cite{Aerts3, Aerts4, Aerts4b} and, more specifically, in his \emph{hidden-measurement approach}~\cite{Aerts4, Aerts4b, Aerts7, Aerts10}.

Aert's explanatory framework~\footnote{Sometimes also referred to as the \emph{Geneva-Brussels approach to the foundations of physics}, as it counts among its founders J. M. Jauch and C. Piron, of Geneva University, and D. Aerts and collaborators, of Brussels Free University; see for instance~\cite{Piron, Piron2, Piron3, Aerts3, Aerts4, Aerts4b, Aerts7, Aerts10, Aerts1, Aerts2, Aerts5, Aerts6, Aerts8} and the references cited therein.} has been substantiated, over the years, by a number of amazing machine-models. These are conventional macroscopic mechanical objects, like those we encounter in our everyday life, that, surprisingly, are able to reproduce not only the strange behavior of pure quantum systems, but also the behavior of more general intermediate structures, which are neither quantum nor classical, but truly intermediate. And since the functioning of these machine-models is fully under our eyes, one can today confidently say, in contrast to Feynman's admonition, that much of the quantum mystery has been in fact unveiled.

Among the most important machine examples invented by Aerts, we can cite his ``connected vessels of water'' model~\cite{Aerts2, Aerts5}, which can reproduce EPR non-local correlations and violate Bell's inequalities, and his $\epsilon$-model~\cite{Aerts3}, describing a point particle on which experiments (measurements) are performed in a very particular way, by exploiting the breakability of peculiar elastic bands (more will be said about it later in the article). In this model, $\epsilon$ is a continuous parameter that can be varied from $0$ to $1$. In the $\epsilon = 0$ limit, the system becomes purely classical, with the outcomes of the measurements that are \emph{a priori} determined by the state of the entity. On the other hand, in the $\epsilon = 1$ limit, the system becomes purely quantum, and is structurally equivalent to the spin of a spin-$1/2$ quantum ``particle,'' reproducing the same transition probabilities that are obtained in a typical Stern-Gerlach experiment. And, for the $0<\epsilon <1$ intermediate values, the system exhibits interesting intermediate structures, which cannot be modelized by a classical phase-space or a quantum Hilbert space~\footnote{Very recently, the $\epsilon$-model has also been successfully applied, as a mathematical tool (in conjunction with the notion of \emph{contextual risk}) to model the ambiguity appearing in so-called Ellsberg paradox, in decision theory and experimental economics~\cite{AertsEco1,AertsEco2}.}.

In this paper we want to follow Aerts' great tradition of inventing macroscopic models that are able to reproduce the behavior of quantum entities, and beyond. More precisely, in the spirit of the above mentioned $\epsilon$-model, we will introduce a new machine-model, which also depends on a parameter $k$: an integer that can be varied from $1$ to a given maximal value $K$. 

For $k=K$ (maximal value) the system can be shown to reproduce the transmission and reflection probabilities of a classical scattering process, whereas for $k=1$ (minimal value) it reproduces the transmission and reflection probabilities of a quantum scattering entity interacting with a Dirac $\delta$-function potential. And, for the $1<k<K$ intermediate values, the machine delivers transmission and reflection probabilities which, in general, cannot be classified as classical or quantum, being truly intermediate.

To this end, in Sec.~\ref{creation-discovery}, we start by introducing, in a didactical way, the general explanatory framework of Aerts' \emph{creation-discovery view} and \emph{hidden-measurement approach}, providing the conceptual language that will allow us to understand the origin of the structural difference between classical, quantum and (quantum-like) intermediate systems. 

In Sec.~\ref{scattering}, we use this language to explain the physical content of transmission and reflection probabilities in one-dimensional quantum scattering systems, and in Sec.~\ref{delta-scattering} we explicitly calculate them for the simple case of a delta-function potential. Then, in Secs.~\ref{Dirac quantum machine}, we present in detail the design and functioning of the $\delta$-quantum machine and show that it reproduces the transmission and reflection probabilities associated to a delta-function potential. 

In Sec.~\ref{model}, we generalize the functioning of the $\delta$-quantum machine in what we call the $k$-model, and show that in the $k=1$ and $k=K$ limit situations it reproduces the quantum and classical probabilities, respectively, whereas for intermediate values of $k$ it can describe processes which are neither classical nor quantum, but truly intermediate.

In Sec.~\ref{comparison}, we highlight the main differences between Aerts' $\epsilon$-model -- which we also describe in some details -- and our $k$-model, and in  Sec.~\ref{potentiality} we use the powerful metaphor of the latter to deepen our understanding of the behavior of quantum (and quantum-like) entities, particularly for what concerns their property of being able to switch from actual to potential modes of being. 

This will take us, in Sec.~\ref{weak}, to the introduction of the new concept of \emph{process-actuality} of a property, that we use to define the related concepts of process-existence and process-macroscopic wholeness. Thanks to these definitions, we will be in a position, in Sec.~\ref{non-spatiality}, to give a precise definition of the important notion of \emph{non-spatiality}, which we show is to be understood not as absence of spatiality, but as existence in an intermediary physical space.  

Finally, in Sec.~\ref{conclusion}, we conclude our work by providing some final remarks.

\section{Creations, discoveries and hidden measurements}
\label{creation-discovery}

The conceptual language we present in this section is mainly the result of the work of two physicists: Constantin Piron (particularly for what concerns the concept of \emph{experimental project}, the definition of the \emph{state} of an \emph{entity} and the precise characterization of the so-called \emph{classical prejudice}) and Diederik Aerts (particularly for his deep analysis of the structural differences between classical and quantum systems, in relation to the various changes an entity can undergo in a measurement process, and the corresponding distinction between classical and quantum probabilities, as expressed in his \emph{creation-discovery view} and, more specifically, in his \emph{hidden measurement approach}). 

The subtlety and richness of the concepts presented in this section would require many more pages of explanation and analysis, also of a mathematical nature. However, the rather succinct and intuitive presentation of this section will certainly suffice for the goal of this article, and we refer the interested reader to the papers of Piron and Aerts that we have mentioned in the Introduction. 

\textbf{Entity}. A physicist' investigation starts when he (she) focus his (her) attention on some specific phenomena, happening in his (her) reality, neglecting some others. To these ensembles of phenomena, which emerge from the others, he (she) can give specific names, and attach properties. In other words, a scientist investigating reality will use his (her) analytical skills to conceptually separate parts of reality having specific sets of properties. These parts are called \emph{entities}. An entity is not necessarily a spatial phenomenon, as it can also refer to mathematical, mental, conceptual aspects of our reality, and many other as well. In other terms, an entity is just an element (not necessarily elementary) of our total reality to which, in our role of participative observers, we are able to attribute specific properties.

\textbf{Property}. Generally speaking, a property is something an entity has independently of the type of context it is confronted with. Properties can either be \emph{actual} or \emph{potential}. If they are actual, it means that the outcomes of those \emph{tests} which are used to (operationally) define them can be predicted, at least in principle, with certainty. On the other hand, if they are potential, it means that such outcomes cannot be predicted with certainty, not even in principle.

\textbf{Experimental project}. The tests that are used to operationally define an entity's properties (and deduce their actuality or potentiality) are \emph{experimental projects} whose outcomes lead to a well-defined ``yes-no'' alternative. They require the specification of: the measuring apparatus to be used, the operations to be performed, and the rule to be applied to unambiguously interpret the results of the experiment in terms of the (mutually excluding) ``yes'' (successful) and ``no'' (unsuccessful) alternatives.

\textbf{State}. By definition, the \emph{state} of an entity is the set of all its actual properties, i.e., the collection of all properties that are actual for an entity in a given moment. And since with time some actual properties become potential, whereas some other potential properties become actual, this means that the state of an entity, in general, changes (i.e., it evolves). In other words, what one can state about an entity in a given moment is different from what one can state about the same entity in the following moment. However, not all properties of an entity will change with time: some of them, usually called \emph{intrinsic properties}, or \emph{attributes}, are more stable, and are usually used to characterize the entity's \emph{identity}, and when they cease to be actual one says that the entity has been destroyed (or partially destroyed).  

\textbf{Classical prejudice}. Being the state the collection of all properties that are actual in a given moment, it's clear that once we know the state of an entity we know, by definition, all it can be said with certainty about it, in that moment. This may lead one to believe that, accordingly, the outcome of whatever test we can perform on the entity is in principle predictable with certainty. Such an AUUA (\emph{Additional Unconsciously Used Assumption}, as Aerts likes to call them; see~\cite{Aerts2}) is usually referred to as the \emph{classical prejudice}: a preconceived idea that was long believed by physicists, but in the end has been falsified by the quantum revolution. 

\textbf{Lack of knowledge}. In the description of an entity, we have to distinguish two kinds of \emph{lack of knowledge}. The first kind is related to our possible incomplete knowledge of the state of the entity, whereas the second kind, much more subtle, is related to our ignorance about the specific interactions arising between the entity and its context, and in particular the experimental testing apparatus.

\textbf{Classical and quantum probabilities}. Every time a scientist is in a situation of lack of knowledge, the best he can do is to formulate probabilistic predictions about the outcome of his experimental projects. Different typologies of lack of knowledge will produce different probabilities. \emph{Classical probabilities} (obeying Kolmogorov's axioms) correspond to situations where the lack of knowledge is only about the state of the entity. Quantum probabilities (not obeying Kolmogorov's axioms) correspond to situations where there is a full knowledge of the state of the entity, but a maximum lack of knowledge about the interaction between the measurement apparatus and the entity. In between these two extremes, one finds intermediate pictures, giving rise to intermediate probabilities which can be neither fitted into a quantum probability model, nor into a classical probability model.     

\textbf{Hidden measurements}. The origin of the structural differences between quantum and classical entities can be more easily understood by introducing the important concept of \emph{hidden measurement} (by measurement we mean here an experiment testing a specific property, or set of properties). In general, measurements are not just observations without effects, as they can provoke real changes in the state of the entity. However, as we usually lack knowledge about the reality of what exactly happens during a measurement process, its outcomes can only be predicted in probabilistic terms. This can be modelized by assuming that to a given (indeterministic) measurement are associated a collection of ``hidden'' \emph{deterministic} measurements, and that when the measurement is performed (on an entity in a given state), one of these hidden measurements does actually take place. In other terms, quantum (or quantum-like~\footnote{The term ``quantum-like'' refers here to those structures that are neither purely classical nor purely quantum, but constitute a more general intermediate picture.}) probabilities find their origin in our lack of knowledge about which one of these hidden (deterministic) measurements effectively take place. 

\textbf{Creation and discovery}. According to the above, classical probabilities express our lack of knowledge about the state of an entity, i.e., about the properties that are already present (i.e., actual) before doing or even deciding doing an experiment. In other words, classical probabilities are about our possibility to \emph{discover} something that is already there. Quantum (or quantum-like) probabilities, on the other hand, express our lack of knowledge about properties that do not exist before the experiment (i.e., are only potential), but are literally \emph{created} (i.e., actualized) by means of the experiment. In other terms, the distinction between classical and quantum probabilities would be just a distinction between discovering what is already there and creating what is still not there, by means of an experiment (i.e., a measurement process). 

\textbf{Soft and hard acts of creations}. We conclude this telegraphic presentation by also mentioning the distinction between \emph{soft} and \emph{hard} creations, as considered by Coecke~\cite{Coecke}. A \emph{soft creation} is a (unitary) structure preserving process that doesn't alter the set of states of an entity, but only the relative actuality and potentiality of its set of properties. On the other hand, a \emph{hard creation} is a process that has the power to alter the set of states of an entity. However, considering that an entity can also be defined by its attribute of having a given set of states, we can say that a hard act of creation is a process that destroys (or partially destroys) the original entity's identity, which therefore disappears from our sight, and creates new entities, suddenly appearing to our sight (or the ``sight'' of our instruments). \emph{N.B.}: a soft act of creation can also be understood as a composite process constituted by a succession of hard acts of creation, whose overall effect results in the restoration of the entity's original identity.

\section{transmissibility and reflectivity}
\label{scattering}

Using the conceptual language we have introduced in the previous section, we shall now describe a typical one-dimensional quantum scattering process and the associated transmission and reflection probabilities~\footnote{One-dimensional systems arise for instance in the study of semiconductor heterostructures and their transport properties. Indeed, in nearly all the situations of physical interest, the device dimensions perpendicular to the transport $x$-direction are greater than the $x$-extension of the potential, thus allowing a separation of the three-dimensional Schr\"odinger equation into a two-dimensional free-electron motion and a one dimensional effective problem~\cite{Vassel}.}.

A quantum entity, like an electron, is an entity characterized by some intrinsic properties, like its spin, its charge and its mass; these are attributes that will remain constantly actual, for as long as the entity exists (i.e., is not destroyed). In addition to them, the state of the quantum entity is characterized by a number of non-intrinsic properties, whose actuality and potentiality may vary as time passes by, like for instance the property of ``being present in a given region of space,'' ``having the momentum in a given cone,'' ``having the spin oriented in a given spatial direction,'' and so on.

When the entity interacts with a force field, described by a potential function $V(x)$, it can typically be in a bound state, and therefore remains localized in the interaction region for all times, or be in a scattering state, propagating away from any bounded region, as $t\to\pm\infty$~\footnote{Here there is of course an abuse of language: since quantum entities are in general non-spatial~\cite{Massimiliano}, instead of saying that the entity ``remains localized in the interaction region for all times,'' we should say that it``has a high probability of being detected in the interaction region for all times.'' Seemingly, instead of saying that it ``propagates away from any bounded region, as $t\to\pm\infty$,'' we should say that it``has a high probability of being detected far away from any bounded region, as $t\to\pm\infty$.''}. 

Let us assume that the quantum entity under consideration is, at time $t$, in a scattering state $|\psi_t\rangle\in {\cal H}$, which we assume to be one-dimensional, i.e., ${\cal H}=L^2(\mathbb{R})$. 

In a typical scattering experiment, one makes sure to prepare the entity, in the remote past, in a suitable free evolving state $|\varphi_t\rangle = e^{-\frac{i}{\hbar}H_0t}|\varphi\rangle$, where $H_0 = -\frac{\hbar^2}{2m}\partial_x^2$ is the free Hamiltonian. This can be expressed by the asymptotic condition: $|\psi_t\rangle \approx |\varphi_t\rangle$, as $t\to-\infty$, where the symbol ``$\approx$'' denotes that the difference between the left and right sides tends to zero. 

Let us assume that the entity approaches the interaction region from the left. This means it has been prepared in the past in a state of positive momentum. More precisely, if $P_+=|+\rangle\langle +|$ is the projection operator onto the set of states of positive momentum, i.e., the set of states actualizing the property of ``having a positive momentum,'' then approaching the potential from the left means that $P_+|\varphi\rangle = |\varphi\rangle$. 

A one-dimensional scattering process can be used as an experimental project to test two specific properties of the quantum entity: \emph{transmissibility} and \emph{reflectivity} [more exactly, one should speak of $(V,\varphi)$-transmissibility and $(V,\varphi)$-reflectivity, as transmission and reflection are defined only in relation to a specific potential $V$ and initial state $|\varphi\rangle$]. 

For the transmissibility (resp. reflectivity) test, the operations to be performed are the following: observe, by means of a suitable measuring apparatus (the details of which we don't need to describe here) if the entity which has been duly prepared in the past, will be detected, in the distant future, in the right side (resp. left side) of the potential region, far away from it. If the apparatus reveals the presence of the entity in the mentioned spatial region, as $t\to\infty$, the test is considered successful (``yes'' answer) and the property of transmissibility (resp. reflectivity) will be said to have been confirmed.

Let us limit our considerations to the transmission case (the reflection one, \emph{mutatis mutandis}, being similar). The property ``being present in the far right of the potential'' can be associated in QM to the projection operator $P_b=\int_b^\infty dx |x\rangle\langle x|$, onto the set of states localized in the spatial interval $(b,\infty)$, where $b$ is a positive large number. Accordingly, transmissibility can be defined as the property for the scattering entity of ``being present in the far right of the potential in the distant future,'' i.e., the property that $|\psi_t\rangle\in P_b{\cal H}$, as $t\to\infty$. 

If the incoming entity is a classical particle, by knowing its state we can easily predict in advance, with certainty, the outcome of the above transmissibility test, in accordance with the classical prejudice mentioned in the previous section. Indeed, a classical particle will be transmitted if and only if its incoming energy $E$ is strictly above the potential, i.e., iff $E>\sup_x V(x)$. 

On the other hand, in QM a full knowledge of the entity's state (at whatever moment) is not sufficient to pre-determine the outcome of the transmissibility test. In other terms, transmissibility remains a potential (uncreated) property, for as long as the test is not performed, and can only possibly be actualized (created) if one effectively performs it.

As any student of QM learns, the best one can do is to predict the outcome of the transmissibility test in probabilistic terms, using for this the \emph{Born rule} (which was formulated by Born in a 1926 paper~\cite{Born}, precisely in the context of a scattering problem). More precisely, following the above discussion, the quantum probability for a successful outcome of the transmissibility test (simply called ``transmission probability'') is given by: 
\begin{eqnarray}
\label{transmission probability definition}
P_{\texttt{tr}}(\varphi) &=&\lim_{t\to\infty} \| P_b\psi_t\|^2 =\lim_{t\to\infty} \| P_b e^{-\frac{i}{\hbar}H_0t} S\varphi \|^2\\
\label{transmission probability definition2}
&=& \lim_{t\to\infty}\|P_+e^{-\frac{i}{\hbar}H_0t}S\varphi\|^2 = \|P_+S\varphi\|^2.
\end{eqnarray}

The second equality in (\ref{transmission probability definition}) follows from the future asymptotic condition $|\psi_t\rangle\approx e^{-\frac{i}{\hbar}H_0t} S|\varphi\rangle$, as $t\to\infty$ ($S$ being the unitary scattering operator), whereas the first equality in (\ref{transmission probability definition2}) expresses the intuitively evident fact that the probability of finding the scattering entity in the region $(b,\infty)$, as $t\to\infty$, is the same as the probability for the entity to propagate in the direction in which it will eventually penetrate into that region, which in the present case corresponds to the probability of having positive momentum (mathematically, this fact follows from the well known Dollard's Scattering-Into-Cones formula; see for instance~\cite{Amrein}). Finally, the last equality in (\ref{transmission probability definition2}) simply follows from the fact that $[H_0,P_+]=0$.

At this point, considering that the scattering operator $S$ commutes with the free Hamiltonian and that $|\varphi\rangle = P_+ |\varphi\rangle$, by defining the transmission operator $T=\langle +|S|+\rangle$, we can write the transmission probability (\ref{transmission probability definition2}) in the form: 
\begin{equation}
\label{transmission probability3}
P_{\texttt{tr}}(\varphi)=\int_0^\infty \!\! dE\, |T(E)|^2 |\varphi(E)|^2.
\end{equation}
Therefore, assuming that the incoming entity from the left has been also prepared in order to actualize the property of ``having a well defined energy,'' meaning that the incoming wave packet is sharply peaked about, say, energy $E$, so that: 
\begin{equation}
\label{monoenergetic limit}
|\varphi(E^\prime)|^2\approx \delta (E^\prime -E),
\end{equation} 
we obtain that the quantum transmission probability becomes:
\begin{equation}
\label{monoenergetic transmission probability}
P_{\texttt{tr}}(\varphi)\approx |T(E)|^2.
\end{equation} 
Of course, a similar approximation holds for the reflection case, yielding for the quantum reflection probability:
\begin{equation}
\label{monoenergetic reflection probability}
P_{\texttt{re}}(\varphi)\approx |R(E)|^2 = 1- |T(E)|^2,
\end{equation} 
where $R(E)$ is the on-shell element of the reflection (from the left) operator $R=\langle -|S|+\rangle$, at energy $E$.

\section{The Dirac $\delta$-function potential}
\label{delta-scattering}

Having clarified the meaning of transmissibility and reflectivity in a quantum scattering process, and derived the corresponding transmission and reflection probabilities, we want now to explicitly calculate them in the simple case of a delta-function potential: $V(x)=\lambda\delta(x)$. This can be easily done by directly solving the stationary Schr\"odinger equation:
\begin{equation}
\label{stationary Schroedinger equation}
\left[-\frac{\hbar^2}{2m}\frac{\partial^2}{\partial x^2}+ \lambda\delta(x)\right]\psi(E,x)=E\psi(E,x),
\end{equation}
with boundary conditions (describing an entity coming from the left)
\begin{equation}
\label{asymptotic condition from the left}
\psi(E,x)=
\begin{cases}
e^{ikx}+R(E) e^{-ikx}, & x<0\\
T(E) e^{ikx}, & x>0,
\end{cases}
\end{equation}
where $R(E)$ and $T(E)$ are the reflection and transmission amplitudes, at energy $E=\hbar^2k^2/2m$. Continuity of $\psi(E,x)$ at $x=0$, yields:
\begin{equation}
\label{equality1}
1 + R(E) = T(E).
\end{equation}
Integrating (\ref{stationary Schroedinger equation}) from $-\epsilon$ to $\epsilon$, using the properties of the delta-function, then taking the limit $\epsilon\to 0$, one obtains the second equality:
\begin{equation}
\label{equality2}
ik\left[R(E)-1\right]=\left(\frac{2m\lambda}{\hbar^2} - ik\right)T(E).
\end{equation}
By combining (\ref{equality1}) and (\ref{equality2}), one then obtains:
\begin{equation}
\label{transmission and reflection amplitudes}
T(E) = \frac{i\kappa}{i\kappa-1}, \quad\quad R(E) = \frac{1}{i\kappa-1},
\end{equation}
where we have defined $\kappa = \hbar^2k/m\lambda = \sqrt{2\hbar^2E/m}/\lambda$. 

Finally, taking the square modulus of the above amplitudes, one gets the quantum transmission and reflection probabilities at fixed energy $E$: 
\begin{equation}
\label{transmission probability}
P_{\texttt{tr}}(E)=|T(E)|^2=\frac{\kappa^2}{1+\kappa^2},
\end{equation}
\begin{equation}
\label{reflection probability}
P_{\texttt{re}}(E) =|R(E)|^2= \frac{1}{1+\kappa^2}.
\end{equation}
In the following, to simplify the discussion, we set the coupling $\lambda =\sqrt{2\hbar^2/m}$, so that $\kappa = \sqrt{E}$ and, simply:
\begin{equation}
\label{transmission and reflection probabilities2}
P_{\texttt{tr}}(E)=\frac{E}{1+E}, \quad\quad P_{\texttt{re}}(E) = \frac{1}{1+E}.
\end{equation}

\section{The $\delta$-quantum machine}
\label{Dirac quantum machine}

In this section, we describe the (Dirac) $\delta$-quantum machine and its functioning, and show that it reproduces the quantum transmission and reflection probabilities (\ref{transmission and reflection probabilities2}).

The entity under study, that we shall call $S_K$, is a macroscopic compound object made of $K$ tiny spheres, having all same mass and density, which can either be positively or negatively electrically charged. The spheres are assumed to be able (in normal conditions) to remain in contact together, for instance because they are slightly magnetic, thus forming a whole cluster-entity.

Entity $S_K$ possesses many distinctive attributes that characterize its identity, like the number $K$ of its components, its total mass $M=\mu K$, with $\mu$ the mass of a single sphere, the material it is made of (that we don't need to specify here), and many other as well. And of course, for $S_K$ to continue to exist in our physical space, all these defining attributes have to remain actual. 

But, besides its more stable attributes, $S_K$ can also assume different states. For instance, it can occupy different spatial locations, have different orientations, shapes, and so on. Some of these states are the result of specific preparations, i.e., determinative processes through which specific states for $S_K$ are selected. Others can be the result of \emph{measurement} processes which, contrary to preparations, are in general non-determinative, but only interrogative, so that their outcomes cannot in general be predicted with certainty. 

In the following we are interested in those preparations that correspond to the different electric charges that $S_K$ can support. As we said, we assume that each one of the $K$ constituent spheres can either assume a positive electric charge $q>0$, or a negative charge $-q$, but cannot be electrically neutral. Therefore, $S_K$ can be prepared in $K+1$ different electric states, each characterized by a specific electric charge: 
\begin{equation}
\label{Q-variable}
Q=q(K_+-K_-)\in\{-qK, -q(K-2),\dots,qK\},
\end{equation}
where $K_+$ and $K_-$ are the number of spheres having positive and negative electric charge, respectively, and $K_++K_-=K$. For instance, an entity $S_5$ can be prepared in 6 different electric states, characterized by the charges: $-5q, -3q, -q, q, 3q, 5q$. 

For later convenience, we also introduce the variable
\begin{equation}
\label{E-variable}
E=\frac{K_+}{K_-}\in\{0, \frac{1}{K-1},\dots,\frac{K-1}{1},\infty\},
\end{equation}
and observe that the charge $Q$ can be entirely expressed in terms of $E$ by the formula $Q=Kq(E-1)/(E+1)$. Hence, we can equivalently parameterize the electric states of $S_K$ by using $E$ instead of $Q$.

\begin{figure}[!ht]
\centering
\includegraphics[scale =.7]{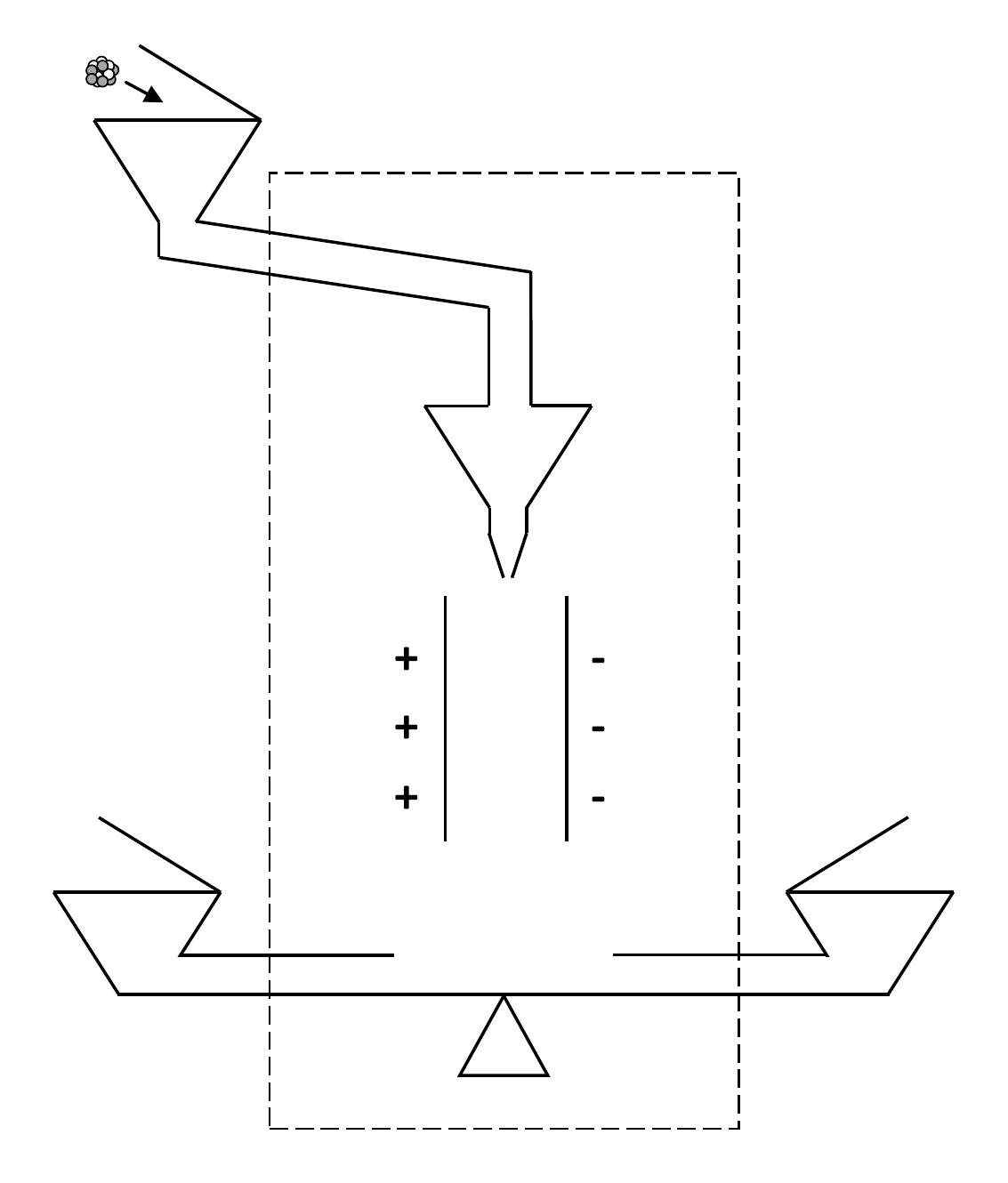}
\caption{A simplified design of the $\delta$-quantum machine.
\label{delta quantum machine}}
\end{figure}

Once we have prepared the entity $S_K$ in a given electric state $E$, we may want to perform some experiments, like for instance a scattering experiment. For this, we use a specific experimental apparatus, consisting of a box with a left upper entry compartment, and two left and right lower exit compartments (see Figure~\ref{delta quantum machine}). 

The box contains some mysterious mechanism that causes the entity which is introduced inside the left upper compartment to either exit in the right lower compartment or in the left one. And we shall say that the entity has the property of \emph{transmissibility} (resp. \emph{reflectivity}) if, once introduced in the box, it ends its run in the right (resp. left) lower exit compartment, with certainty.  

More precisely, the experimental protocol is as follows: prepare entity $S_K$ in a given state $E$ and place it in the entry compartment, wait until the machine stops producing noise, then look in the two lower exit compartments. If you find $S_K$ in the right one, the outcome of the experiment is a ``yes'' if the experiment is used to test transmissibility, and ``no'' if it is used to test reflectivity. Conversely, if you find $S_K$ in the left one, the outcome is ``no'' if the experiment is used to test transmissibility, and ``yes'' if it is used to test reflectivity. 

Now, the mechanism inside the box, that we are going to describe below, is such that if one performs for each incoming state $E$ a number of experiments, then calculate the relative frequencies of transmitted and reflected events, one finds that these relative frequencies converge (as the number of experiments increases) to the quantum transmission and reflection probabilities (\ref{transmission and reflection probabilities2}). 

In other terms, entity $S_K$ and the measuring apparatus constitute a Dirac \emph{$\delta$-quantum machine}, in the sense that the system is isomorphic (from a probabilistic point of view) to the one-dimensional scattering of a quantum entity by a delta-function potential.

Let us now reveal the mystery and describe the interior of the machine (see Figure~\ref{delta quantum machine}). Once the entity has been introduced in the entry compartment, it rolls down along a tube and falls inside a central internal compartment. Due to the impact against the walls of the compartment, the composite entity (which is quite fragile, as the magnetic cohesion of the spheres is low) breaks into its $K$ components. 

Then, a specific shutter mechanism selects a single sphere and lets it fall exactly in the middle of the two charged plates of a parallel-plate capacitor (condenser). Assuming that the left and right plates are positively and negatively charged, respectively, as it falls the sphere is deviated to the right if its charge is $q>0$, or to the left if its charge is $-q$. And, at the end of its fall, it lands on an horizontal lever which is in equilibrium on its pivot (like a seesaw). 

If the falling particle is positively charged, its landing point will be on the right side of the pivot and therefore its weight will cause the lever (which makes a single whole with the two exit compartments) to go down to the right. In this way, the sphere will reach the right exit compartment and remain there, causing the lever to maintain its inclination to the right (or to the left, for a negatively charged sphere). 

Then, after a little while, the automatic shutter mechanism frees another sphere, which can either be positively or negatively charged. If it is positive, it will fall to the right and reach the first positive sphere inside the right exit compartment. On the other hand, if it is negative, the capacitor electric field will cause it to deviate to the left and land on the left side of the lever pivot. However, since the first sphere already reached the far right of the lever, its torque (moment of force) will not be sufficient to turn the lever to the left. Thus, following a brief exploration of part of the left hand side of the lever, it will revert its motion and also end its run inside the right compartment. 

The process continues in this way, with the shutter mechanism releasing one sphere after the other (only one sphere at a time passes through the capacitor), with all of them ending their journey either inside the right compartment (if the first sphere was positive) or inside the left one (if the first sphere was negative), rebuilding in this way the whole composite entity $S_K$.

Now that we have described the internal working of the box, which therefore is no longer a mystery, we are in a position to calculate the transmission and reflection probabilities. The calculation is very simple, as the transmission probability is nothing but the probability that the first sphere selected by the shutter mechanism is positive, which is given by the ratio:
\begin{equation}
\label{ratio1}
P_{\texttt{tr}}(E)=\frac{K_+}{K}=\frac{K_+}{K_-+K_+}=\frac{E}{1+E}.
\end{equation}
Similarly, the reflection probability is given by the ratio 
\begin{equation}
\label{ratio2}
P_{\texttt{re}}(E)=\frac{K_-}{K}=\frac{K_-}{K_-+K_+}=\frac{1}{1+E}.
\end{equation}

In other terms, the macroscopic $\delta$-quantum machine exactly reproduces the quantum transmission and reflection probabilities (\ref{transmission and reflection probabilities2}). (Of course, the machine cannot reproduce every scattering process associated to every incoming energy $E$ in the interval $E\in [0,\infty)$, but only a finite subset of them, precisely those for which the incoming energy $E$ is of the form $K_+/K_-$).  

Let us now analyze the functioning of the $\delta$-quantum machine using the language of Aert's hidden measurement approach. First of all, we can observe that the only circumstances in which we can predict in advance if $S_K$ will be transmitted or reflected are the $E=0$ and $E=\infty$ situations, which are the analogues of the low and high-energy limits in potential scattering. These correspond to an entity $S_K$ which has been prepared in electric states only composed of negatively or positively charged spheres, so that the first sphere selected by the shutter mechanism will certainly be negative or positive. 

In other words, for the $E=0$ preparation the reflectivity property is actual for $S_K$, whereas it is the transmissibility property to be actual for the $E=\infty$ preparation. 

On the other hand, for states such that $E\neq 0,\infty$, it is impossible to predict if the particle will be transmitted or reflected. This because it is impossible to predict which sphere will be selected first, as the experimenter cannot control the way in which $S_K$ rolls down in the inlet tube and disassemble by colliding against the walls of the internal compartment, thus causing one of the spheres to be selected by the shutter mechanism.   

In other terms, if the composing spheres are not all positively or negatively charged, each time we introduce the entity in the entry compartment, we cannot know in advance which one of its composing spheres will be chosen first by the shutter mechanism.  

This means that the measurement involves an element of choice between $K$ ``hidden'' measurements, each one corresponding to a different possible selection for the first sphere by the shutter mechanism. And since these $K$ choices are not under the control of the experimenter, this is the reason why we can say that the measurement process involves hidden measurements, and that it is the lack of knowledge about which individual deterministic measurements is selected that is responsible for our inability to predict in advance the outcome of the experiment, but only evaluate it in probabilistic terms.

\section{The $k$-model}
\label{model}
 
In this section we want to enlarge the class of measurements that we can perform on entity $S_K$, and will do so by modifying the functioning of the shutter mechanism in the box. This will allow us to derive transmission and reflection probabilities which cannot be described neither by a classical nor by a quantum scattering system. 

Given an entity $S_K$, with a specific electric state $E$, we consider now a set of $K$ structurally different measurements, indexed by an integer number $k\in\{1,\dots,K\}$ (not to be confused with the momentum defined in Sec.~\ref{delta-scattering}). In a $k$-measurement, the shutter mechanism selects each time $k$ spheres simultaneously, and let them fall, at the same time, in the charged plates of the capacitor. 

If the majority of the $k$ selected spheres are positively (resp. negatively) charged, the number of them which will be deviated to the right (resp. left) will be greater than the number that will be deviated to the left (resp. right), thus causing the lever to go down to the right (resp. left), so that all $k$ spheres will finally end their run inside the right (resp. left) compartment. 

The process then continues, with the shutter mechanism selecting another tranche of $k$ spheres (or less, if there are not enough left), and will do so until all spheres will have been released, in successive tranches, in the capacitor. 

Clearly, for the same reasons we have explained in the $k=1$ analysis of the previous section, once the first tranche of $k$ spheres has caused the lever to slant right (resp. left), the following tranches of $k$ (or less) spheres will not be able to subsequently change the inclination of the lever, so that all $K$ spheres will in the end reassemble in the right (resp. left) compartment, thus recreating the whole entity $S_K$.

But what if $k$ is even and the first selected tranche of $k$ spheres contains exactly $k/2$ positive and $k/2$ negative spheres? In this case, a same amount of spheres will be deviated on the left and on the right by the capacitor's electric field. Thus, an equal number of them will land on the left and right sides of the pivot. This is clearly a symmetric situation. However, it is also an unstable one, and the slightest fluctuation in the system will finally break the symmetry and cause the lever to tilt either to the left or to the right. And because nothing favor the left or right tilting, we can associate a probability $1/2$ to the two outcomes.

Let us now calculate the transmission probability $P_{\texttt{tr}}^{k}(E)$ for a $k$-measurement, $1\leq k\leq K$, and an entity $S_K$ prepared in an electric state $E=K_+/K_-$. The calculation is quite simple, as the transmission probability is nothing but the probability that the total electric charge of the first tranche of $k$ selected spheres is strictly positive, plus $1/2$ times the probability that the charge is zero. 

Using the \emph{binomial coefficient} and considering first the case where $k$ is an odd integer, i.e., $k=1,3,5,\dots$ (meaning that the zero charge circumstance cannot arise), we have: 
\begin{equation}
\label{n-odd}
P_{\texttt{tr}}^{k}(E)=\sum_{m=0}^{\frac{k-1}{2}}\frac{\binom {K_+}{k-m}\binom {K_-}m}{\binom Kk}.
\end{equation}
On the other hand, for $k$ even, i.e., $k=2,4,6,\dots$, we have the weighted formula: 
\begin{equation}
\label{n-even}
P_{\texttt{tr}}^{k}(E)=\sum_{m=0}^{\frac{k-2}{2}}\frac{\binom {K_+}{k-m}\binom {K_-}m}{\binom Kk}+\frac{1}{2}\frac{\binom {K_+}{\frac{k}{2}}\binom {K_-}{\frac{k}{2}}}{\binom Kk}.
\end{equation}

These formulae can be explicitly evaluated for different values of $k$. For instance, a straightforward calculation yields for $k=1,2$:
\begin{equation}
\label{n=1,2}
P_{\texttt{tr}}^{1}(E)=P_{\texttt{tr}}^{2}(E)=\frac{K_+}{K}.
\end{equation}
For $k=3,4$, one finds:
\begin{equation}
\label{n=3,4}
P_{\texttt{tr}}^{3}(E)=P_{\texttt{tr}}^{4}(E)=\frac{K_+(K_+-1)(3K-2K_+ -2)}{K(K-1)(K-2)}.
\end{equation}

Longer explicit formulae can easily be written for higher values of $k$. In general, one finds that probabilities for increasing $k$ are pair wise equal, i.e., $P_{\texttt{tr}}^{k}(E)=P_{\texttt{tr}}^{k+1}(E)$, for $k=1,3,5,\dots$ (a combinatorial fact that we shall not prove in this article, as it has no particular relevance for our discussion). 

Considering entities $S_3$, $S_5$ and $S_7$, an explicit calculation yields, for the transmission probabilities, the values given in Tables~\ref{table3}, \ref{table5} and \ref{table7}, respectively.
\begin{table}[!ht]
\begin{center}
\begin{tabular}{|c|c|c|c|c|}
\hline
$k\backslash E$ & $0$ & $\frac{1}{2}$ & $\frac{2}{1}$ & $\infty$\\
\hline
$1$ & 0 & $\frac{1}{3}$ & $\frac{2}{3}$ & 1\\
\hline
$2$ & 0 & $\frac{1}{3}$ & $\frac{2}{3}$ & 1\\
\hline
$3$ & 0 & 0 & 1 & 1\\
\hline
\end{tabular}
\caption{\label{table3}The transmission probabilities $P_{\texttt{tr}}^{k}(E$) for a compound entity $S_3$, made of $3$ spheres.}
\end{center}
\end{table}
\begin{table}[!ht]
\begin{center}
\begin{tabular}{|c|c|c|c|c|c|c|}
\hline
$k\backslash E$ & $0$ & $\frac{1}{4}$ & $\frac{2}{3}$ & $\frac{3}{2}$& $\frac{4}{1}$ & $\infty$\\
\hline
$1$ & 0 & $\frac{1}{5}$ & $\frac{2}{5}$ & $\frac{3}{5}$ & $\frac{4}{5}$ & 1\\
\hline
$2$ & 0 & $\frac{1}{5}$ & $\frac{2}{5}$ & $\frac{3}{5}$ & $\frac{4}{5}$ & 1\\
\hline
$3$ & 0 & 0 & $\frac{1.5}{5}$ & $\frac{3.5}{5}$ & 1 & 1\\
\hline
$4$ & 0 & 0 & $\frac{1.5}{5}$ & $\frac{3.5}{5}$ & 1 & 1\\
\hline
$5$ & 0 & 0 & 0 & 1 & 1 & 1\\
\hline
\end{tabular}
\caption{\label{table5}The transmission probabilities $P_{\texttt{tr}}^{k}(E$) for a compound entity $S_5$, made of $5$ spheres.}
\end{center}
\end{table}
\begin{table}[!ht]
\begin{center}
\begin{tabular}{|c|c|c|c|c|c|c|c|c|}
\hline
$k\backslash E$ & $0$ & $\frac{1}{6}$ & $\frac{2}{5}$ & $\frac{3}{4}$& $\frac{4}{3}$ & $\frac{5}{2}$ & $\frac{6}{1}$ & $\infty$\\
\hline
$1$ & 0 & $\frac{1}{7}$ & $\frac{2}{7}$ & $\frac{3}{7}$ & $\frac{4}{7}$ & $\frac{5}{7}$ & $\frac{6}{7}$ & 1\\
\hline
$2$ & 0 & $\frac{1}{7}$ & $\frac{2}{7}$ & $\frac{3}{7}$ & $\frac{4}{7}$ & $\frac{5}{7}$ & $\frac{6}{7}$ & 1\\
\hline
$3$ & 0 & 0 & $\frac{1}{7}$ & $\frac{2.6}{7}$ & $\frac{4.4}{7}$ & $\frac{6}{7}$ & 1 & 1\\
\hline
$4$ & 0 & 0 & $\frac{1}{7}$ & $\frac{2.6}{7}$ & $\frac{4.4}{7}$ & $\frac{6}{7}$ & 1 & 1\\
\hline
$5$ & 0 & 0 & 0 & $\frac{2}{7}$ & $\frac{5}{7}$ & 1 & 1 & 1\\
\hline
$6$ & 0 & 0 & 0 & $\frac{2}{7}$ & $\frac{5}{7}$ & 1 & 1 & 1\\
\hline
$7$ & 0 & 0 & 0 & 0 & 1 & 1 & 1 & 1\\
\hline
\end{tabular}
\caption{\label{table7}The transmission probabilities $P_{\texttt{tr}}^{k}(E$) for a compound entity $S_7$, made of $7$ spheres.}
\end{center}
\end{table}

The $S_3$ case is not particularly interesting, as it doesn't go beyond the quantum and classical structures. Indeed, the $k=1$ and $k=2$ measurements reproduce the pure quantum probabilities (\ref{transmission and reflection probabilities2}), whereas the $k=3$ measurement is isomorphic to a classical scattering experiment, where for each value of $E$ the particle is deterministically either transmitted or reflected.

On the other hand, the case of entity $S_5$ is already rich enough to show new genuine intermediate structures. Again, the $k=1$ and $k=2$ measurements just yield the pure quantum probabilities (\ref{transmission and reflection probabilities2}), and the $k=5$ measurement delivers a pure classical result. On the other hand, the transmission probabilities obtained in the $k=3$ and $k=4$ experiments cannot be associated to a classical or to a quantum scattering experiment.

They cannot be associated to a classical experiment as for a classical particle the outcome of the transmissibility (or reflectivity) test is predetermined, since it can be predicted with certainty. This however is contradicted by the transmission probabilities for $E=2/3,3/2$, whose values are different from $0$ or $1$. 

To see that they cannot be associated to a quantum experiment, i.e., be obtained by solving a one-dimensional stationary Schr\"odinger equation for a given potential $V(x)$, we can use the property of the \emph{Wronskian}~\footnote{We recall that the Wronskian $W(f,g)$ of two functions $f$ and $g$ is given by: $W(f,g)=fg' -gf'$.}. 

It is well known that the Wronskian of two linearly independent solutions of the one-dimensional Schr\"odinger equation is a constant different from zero. Calculating the Wronskian of the two solutions describing an entity coming from the left and from the right, respectively, which are linearly independent for $E>0$, one finds it is proportional to $T(E)$. And since the Wronskian of two linearly independent solutions must be different from zero, we necessarily have that $T(E)\neq 0$, for $E> 0$. 

In other words, apart from possible zero-energy solutions (which, as is the case for the discrete spectrum, are non-degenerate), we find that the one-dimensional Schr\"odinger equation cannot produce zero transmission probabilities above the zero-energy threshold. 

Therefore, the zero transmission probability obtained for $E=1/4$ in the $k=3$ and $k=4$ measurements, cannot be the result of a quantum scattering process. In other terms, we find that the $k=3$ and $k=4$ measurements in the $K=5$ case, are neither classical nor quantum scattering processes, but truly hybrid, intermediate processes.
 
Of course, the existence of intermediate -- neither classical nor quantum -- regimes becomes more and more evident as $K$ increases. Considering for instance the transmission probabilities for $S_7$, given in table~\ref{table7}, we observe that the process is purely quantum for $k=1,2$, intermediate for $k=3,4,5,6$, and classical for $k=7$. And, for a general entity $S_K$, $K\geq 5$, $K$ odd, we have that the regime will be quantum for the $k=1,2$ measurements, intermediate (i.e., quantum-like) for $3\leq k\leq K-1$, and classical for $k=K$. 

The attentive reader will have noticed that we have only considered so far the cases where $K$ is odd. The situation for $K$ even is a little different. The explicit probabilities for $S_2$, $S_4$ and $S_6$ are given in tables~\ref{table2}, \ref{table4} and \ref{table6}, respectively. 
\begin{table}[!ht]
\begin{center}
\begin{tabular}{|c|c|c|c|}
\hline
$k\backslash E$ & $0$ & $\frac{1}{1}$ &  $\infty$\\
\hline
$1$ & 0 & $\frac{1}{2}$  & 1\\
\hline
$2$ & 0 & $\frac{1}{2}$  & 1\\
\hline
\end{tabular}
\caption{\label{table2}The transmission probabilities $P_{\texttt{tr}}^{k}(E$) for a compound entity $S_2$, made of $2$ spheres.}
\end{center}
\end{table}
\begin{table}[!ht]
\begin{center}
\begin{tabular}{|c|c|c|c|c|c|}
\hline
$k\backslash E$ & $0$ & $\frac{1}{3}$ & $\frac{2}{2}$ & $\frac{3}{1}$ & $\infty$\\
\hline
$1$ & 0 & $\frac{1}{4}$ & $\frac{2}{4}$ & $\frac{3}{4}$ & 1\\
\hline
$2$ & 0 & $\frac{1}{4}$ & $\frac{2}{4}$ & $\frac{3}{4}$ & 1\\
\hline
$3$ & 0 & 0 & $\frac{1}{2}$ & 1 & 1\\
\hline
$4$ & 0 & 0 & $\frac{1}{2}$ & 1 & 1\\
\hline
\end{tabular}
\caption{\label{table4}The transmission probabilities $P_{\texttt{tr}}^{k}(E$) for a compound entity $S_4$, made of $4$ spheres.}
\end{center}
\end{table}
\begin{table}[!ht]
\begin{center}
\begin{tabular}{|c|c|c|c|c|c|c|c|}
\hline
$k\backslash E$ & $0$ & $\frac{1}{5}$ & $\frac{2}{4}$ & $\frac{3}{3}$& $\frac{4}{2}$ & $\frac{5}{1}$ & $\infty$\\
\hline
$1$ & 0 & $\frac{1}{6}$ & $\frac{2}{6}$ & $\frac{3}{6}$ & $\frac{4}{6}$ & $\frac{5}{6}$ & 1\\
\hline
$2$ & 0 & $\frac{1}{6}$ & $\frac{2}{6}$ & $\frac{3}{6}$ & $\frac{4}{6}$ & $\frac{5}{6}$ & 1\\
\hline
$3$ & 0 & 0 & $\frac{1.2}{6}$ & $\frac{3}{6}$ & $\frac{4.8}{6}$ & 1 & 1\\
\hline
$4$ & 0 & 0 & $\frac{1.2}{6}$ & $\frac{3}{6}$ & $\frac{4.8}{6}$ & 1 & 1\\
\hline
$5$ & 0 & 0 & 0 & $\frac{1}{2}$ & 1 & 1 & 1\\
\hline
$6$ & 0 & 0 & 0 & $\frac{1}{2}$ & 1 & 1 & 1\\
\hline
\end{tabular}
\caption{\label{table6}The transmission probabilities $P_{\texttt{tr}}^{k}(E$) for a compound entity $S_6$, made of $6$ spheres.}
\end{center}
\end{table}

We can now observe that, as $k$ increases, we don't reach a strict classical regime, where the transmission probability, as a function of the incoming energy $E$, is either $0$ or $1$. Indeed, for the $E=1$ case (i.e., $K_+=K_-=K/2$), the transmission probability is $1/2$, independently of the $k$-measurement considered. 

However, also in the even case we can say that the $K$-measurement [or $(K-1)$-measurement, as they are identical from a probabilistic point of view] can be understood as a classical process. Indeed, also in a classical system it can happen that the incoming energy $E$ is such that $E=\sup_xV(x)$. In this circumstance, the incoming particle approaching the potential will slow down and stop, right at the point where the potential reaches its maximum (or at the first of these points, if there are more than one). This however corresponds to a situation of unstable equilibrium, which in real systems will be easily destroyed by the slightest perturbation, causing the particle to be finally transmitted or reflected. And, if nothing \emph{a priori} favors one of the two processes, the best one can do is to attach an equal probability of $1/2$ to both of them. 

In other terms, one can also compare the $K$ and $K-1$ measurements, for $K$ even, to classical measurements, provided one assumes that they correspond to a situation such that $\sup_xV(x) =1$, so that a particle with incoming energy $E=1$ will be captured by the potential, in a situation of unstable equilibrium, for a certain amount of time, until a random fluctuation will cause it to escape, either to the left or to the right, with equal probability.

\section{Comparing the $\epsilon$-model and the $k$-model}
\label{comparison}

In the previous section we have analyzed, in some detail, the functioning of what we have called the $k$-model: a structurally more complex system generalizing the $\delta$-quantum machine. As $k$ increases, we have seen that one gradually switches from a situation of \emph{maximum lack of knowledge}, described by a purely quantum process, to a situation of \emph{minimum lack of knowledge}, described by a purely classical process, which is reached for $k=K$, in the case where $K$ is odd, and for $k=K-1,K$, in the case where $K$ is even. 

In other terms, as $k$ increases, we gradually decrease our lack of knowledge about the hidden measurements chosen by the machine. Intuitively, this can be understood by observing that the bigger is the size of the first fragment of $S_K$ selected by the shutter (i.e., the bigger is $k$), the easier it is to predict its total electric charge (which is responsible for the outcome). Indeed, the bigger is the size of the fragment selected and the closer is its charge to the charge $Q$ of the whole entity $S_K$. Therefore, the better can we approximate its value. 

More precisely, our ability to predict the outcome of a given experiment, for an entity prepared in a state $E$, is determined by the ratio of the number of favorable hidden experiments to the total number of hidden experiments, which can be selected in a typical $k$-measurement, as expressed by formulae (\ref{n-odd}) and (\ref{n-even}). 

What is interesting to observe is that in the intermediate (quantum-like) situations, which are neither classical nor quantum, there are states for which, although $E\neq 0,\infty$ (i.e., although the spheres composing $S_K$ are not all of the same charge), we are nevertheless in a position to predict with certainty the outcome. This happens each time that $k\geq 2M+1$, where $M=\inf\, \{K_+,K_-\}$, i.e., each time that $k$ exceeds twice the minimum number of equally charged spheres composing $S_K$.

As we have mentioned in the introduction, our $\delta$-quantum machine and $k$-model have been inspired by Aert's spin-quantum machine and $\epsilon$-model~\cite{Aerts3, Aerts4b, Aerts10}. Let us briefly recall what are the basic elements constituting Aert's spin-quantum machine, whose functioning is isomorphic to the description of the spin of a spin $1/2$ entity. Aerts considers an entity which is a simple point particle localized on the surface of a three-dimensional Euclidean sphere of unit radius, the different possible states of which are the different places the particle can occupy on it. 

\begin{figure}[!ht]
\centering
\includegraphics[scale =.6]{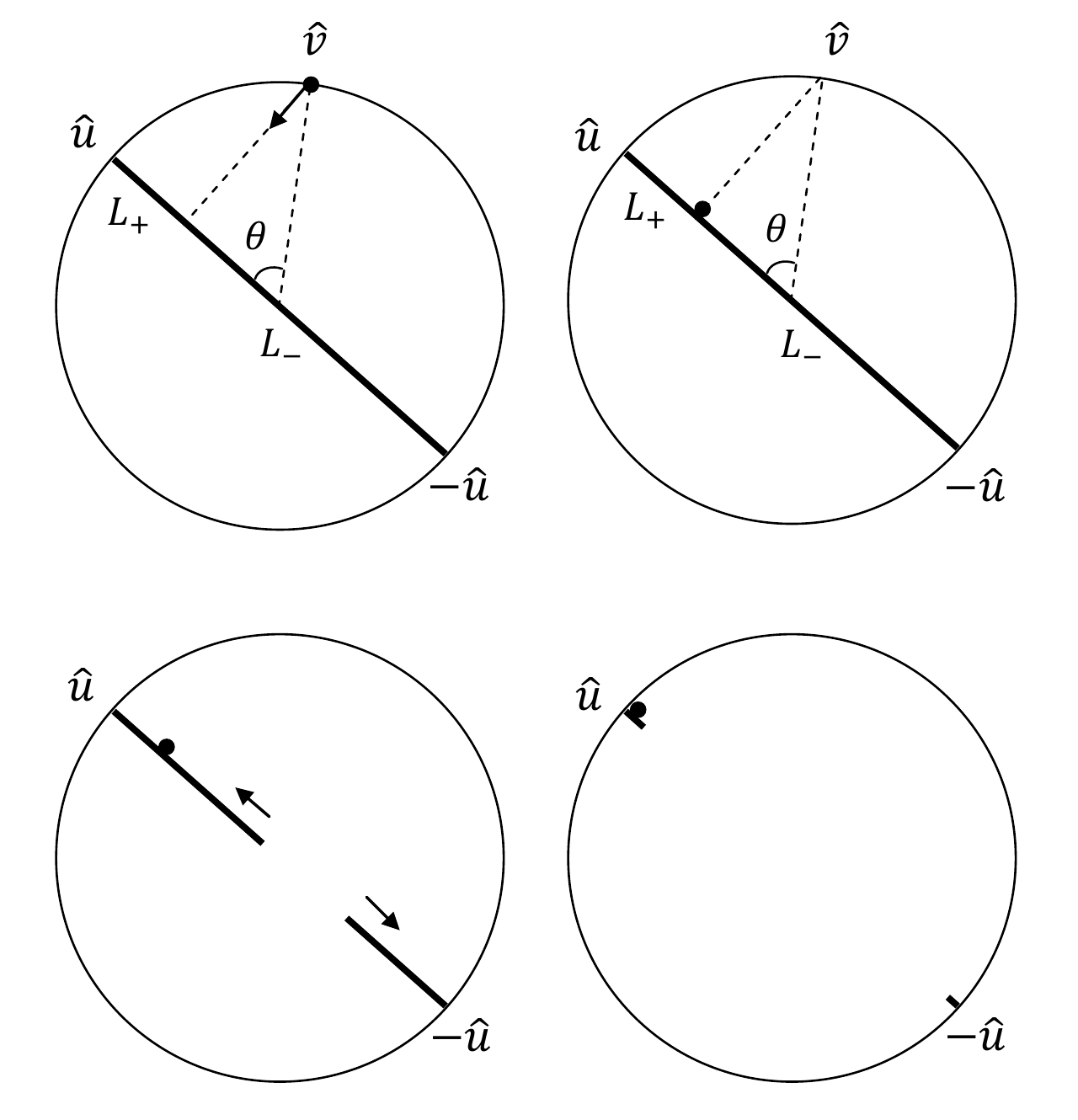}
\caption{A schematic representation of the spin quantum machine measurement process, in the plane of the $3$-dimensional sphere where it takes place.
\label{spin quantum machine}}
\end{figure}

The particularity of the model resides in the way experiments are designed. Indeed, to observe the state of the entity the experimental protocol is to use a sticky elastic band that is stripped between two opposite points of the sphere's surface, identified by two opposite unit vectors $\pm \hat{u}$ (each couple of points defining a different experiment $e_{\hat{u}}$). Then, the procedure is to let the point particle fall from its original location (specified by a unit vector $\hat{v}$) orthogonally onto the elastic and stick to it, then wait until the latter breaks, at some unpredictable point, so that the particle, which is attached to one of the two pieces of it, will be pulled to one of the opposite end points $\pm \hat{u}$, thus producing the outcome of the experiment, i.e., the state that is acquired by the entity as a result of the elastic $e_{\hat{u}}$-measurement (see Fig.~\ref{spin quantum machine}).

It is then straightforward to calculate, with some elementary trigonometry, the probabilities of the different possible outcomes and show that they exactly reproduce those obtained in typical Stern-Gerlach measurements on spin-$1/2$ quantum entities~\cite{Aerts3,Aerts4, Aerts4b}. Indeed, the probability that the particle ends up in point $\pm \hat{u}$ is given by the length $L_\pm$ of the piece of elastic between the particle and the end-point, divided by the total length of the elastic (which is twice the unit radius). Therefore, if $\theta$ is the angle indicated in Figure~\ref{spin quantum machine}, between vectors $\hat{u}$ and $\hat{v}$, we have that the probability for the outcome $\pm\hat{u}$ is given by:
\begin{equation}
\label{probabilities quantum machine}
P_{\pm\hat{u}}(\hat{v}) =\frac{1}{2}(1\pm\cos\theta)=
\begin{cases}
\cos^2\frac{\theta}{2} \\
\sin^2\frac{\theta}{2}, 
\end{cases}
 \end{equation}
which is exactly the quantum probability for measuring the spin of a spin-$1/2$ quantum entity. 

\begin{figure}[!ht]
\centering
\includegraphics[scale =.6]{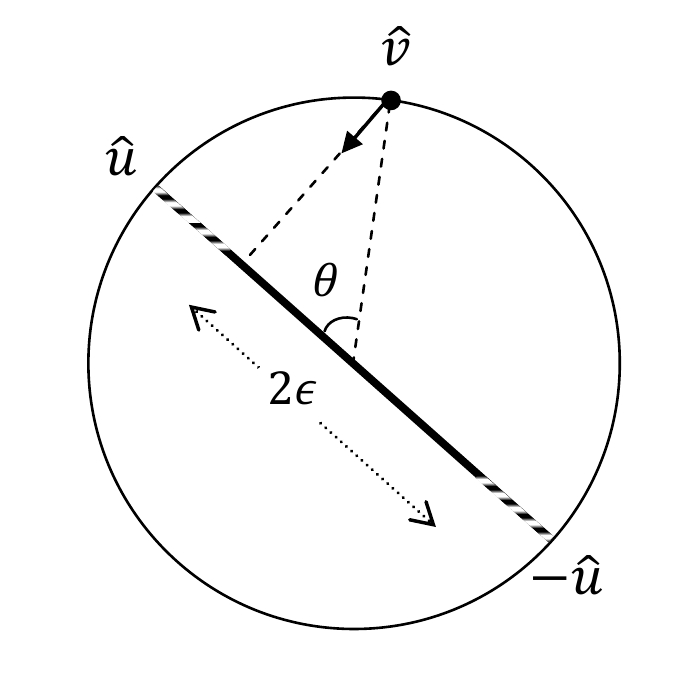}
\caption{A schematic representation of a measurement using an $\epsilon$-elastic, which can only break in its central segment of length $2\epsilon$, whereas it is unbreakable in its lower and upper segments of length $\frac{1}{2}-\epsilon$.
\label{spin epsilon machine}}
\end{figure}

On the basis of his spin-quantum machine, Aerts then considers a more general machine, called the $\epsilon$-model~\cite{Aerts3}, employing elastics of a more complex structure. More precisely, Aerts introduces what he calls $\epsilon$-elastics (we describe here a simplified version of the model, presented in~\cite{Aerts4b}) which are uniformly breakable only in a segment of length $2\epsilon$ around their center, and unbreakable in their lower and upper segments (see Figure~\ref{spin epsilon machine}).

An $\epsilon =1$ measurement (i.e., a measurement using a uniformly breaking $1$-elastic) corresponds to the pure quantum situation with a maximum lack of knowledge about the point where the elastic is going to break. This is the situation of the simple spin-quantum machine that we have previously described, whose probabilities are given by (\ref{probabilities quantum machine}). An $\epsilon =0$ measurement (i.e., a measurement using a $0$-elastic) corresponds to a pure classical situation with minimum lack of knowledge, where the elastic is going to break with certainty in the middle (i.e., in a predetermined point). 

On the other hand, a general $\epsilon$-measurement, with $0<\epsilon <1$, using an $\epsilon$-elastic which can (uniformly) break only around its center, in a segment of length $2\epsilon$, corresponds to a quantum-like situation (which is neither quantum nor classical) of intermediate knowledge.
The associated probabilities are easy to calculate and one has to distinguish the following three cases:

(1) If the particle, when it falls orthogonally onto the elastic, lands on its upper unbreakable segment ($\hat{v}\cdot\hat{u}\geq\epsilon$), then:
\begin{equation}
\label{probabilities epsilon machine1}
P_{\pm\hat{u}}^{\epsilon}(\hat{v})=
\begin{cases}
1 \\
0, 
\end{cases}
 \end{equation}

(2) If the particle, when it falls orthogonally onto the elastic, lands on its central uniformly breakable segment of length $2\epsilon$, ($-\epsilon < \hat{v}\cdot\hat{u}<\epsilon$), then:
\begin{equation}
\label{probabilities epsilon machine2}
P_{\pm\hat{u}}^{\epsilon}(\hat{v})=\frac{1}{2\epsilon}(\epsilon \pm\cos\theta).
\end{equation}

(3) If the particle, when it falls orthogonally onto the elastic, lands on its lower unbreakable segment ($\hat{v}\cdot\hat{u}\leq -\epsilon$), then:
\begin{equation}
\label{probabilities epsilon machine3}
P_{\pm\hat{u}}^{\epsilon}(\hat{v})=
\begin{cases}
0 \\
1. 
\end{cases}
 \end{equation}

Clearly, the $\epsilon$ parameter plays in Aerts's $\epsilon$-model the same role of the $k$-parameter in our $k$-model: by varying it one varies the level of knowledge (or level of control) the experimenter has in relation to the experiment performed, describing in this way a (here continuous) transition from purely quantum ($\epsilon=1$), to quantum-like ($0<\epsilon <1$), to purely classical regimes ($\epsilon=0$); see~\cite{Aerts-Durt} for more details about this transition.

Of course, there are many differences between our $\delta$-quantum machine and corresponding $k$-model, and Aert's spin-quantum machine and corresponding $\epsilon$-model. One is the obvious fact that they modelize different physical systems and therefore yield different probabilities. Another one is the greater structural richness of Aerts' $\epsilon$-model. 

This is so not only because the $\epsilon$ parameter is continuous, whereas the $k$ parameter is discrete, but also because for a given $\epsilon$ and state of the point particle entity (specified by the unit vector $\hat{v}$), there is in Aerts' model an infinity of different possible experiments $e_{\hat{u}}$, corresponding to the different possible orientations of the elastic (specified by the unit vector $\hat{u}$). On the other hand, for a given $k$ and state of $S_K$, there is in our model only a single possible experiment. 

Of course, this is how it should be, seeing that there is only a single spatial direction in a one-dimensional scattering experiment. However, this greater structural richness which is present in Aerts' model becomes essential if one wants to rigorously prove the non-Kolmogorovian nature of the probability model involved, as for this at least three different experiments are needed~\cite{Aerts7,Aerts10,Aerts11}.

On a different level, there is another important difference between the two models: in Aert's quantum machine the ``breaking mechanism,'' which is at the origin of randomness, is associated to the measuring apparatus (the breaking of the elastic band), whereas in our quantum machine the ``breaking mechanism'' is associated to the entity itself, which during the course of the measurement is temporarily disassembled. 

In other terms, in Aert's model the entity under study, a classical point particle, always remains present and localized in our three-dimensional Euclidean space, during the entire experiment. On the other hand, entity $S_K$ of our model is present in our three-dimensional space only at the beginning of the experiment, when it is introduced in the machine, in a given electric state, and at the end of it, when its presence is again observed in one of the two left and right exit compartments.

\section{The potential mode of being}
\label{potentiality}

Following the discussion of the previous section, a natural question arises: What happens to entity $S_K$ during the course of the experiment? As the functioning of the machine presents no mysteries, we can easily answer this question.  For this, we first have to remember that entity $S_K$ exists in our three-dimensional space for as long as it conserves its identity of being a whole, cohesive cluster-entity, made of $K$ charged spheres. Therefore, in the moment the entity disassembles by falling inside the central internal compartment of the machine, it disappears from our sight, i.e., it disappears from our three-dimensional space.   

In other terms, during the measurement process, $S_K$ is temporarily destroyed and, in its place, smaller entities are created. These smaller entities, which are fragments of $S_K$, interact separately with the different elements of the machine, before being all reassembled together, inside one of the two exit compartments. 

This means that the \emph{mode of being} of $S_K$ changes during the different phases of the measuring process. It \emph{actually exists}, in the sense that it is present in our three-dimensional space, when it is introduced in the machine; it \emph{potentially exists}, in the sense that it is no longer present in our three-dimensional space, when it interacts with the different elements of the machine; it comes again into actual existence, in the sense that it re-emerges (or re-immerge) in our three-dimensional space, when its different fragments are brought once again together. 

What is interesting to observe is that during the measurement process the different fragments originating from $S_K$ explore different regions of the three-dimensional space. Some of them, in certain moments, can be found on the lever on both sides of the pivot, so that, before the measurement is terminated, we can say that the potential entity $S_K$ is in a sort of superposition of partially transmitted and reflected components. 

Therefore, entity $S_K$, while it is in its p\emph{otential mode of being}, which is a \emph{non-spatial} mode of being (relative to our three-dimensional space), can behave as a genuine \emph{non-local} entity, i.e., an entity made of parts which, seen from our ordinary three-dimensional perspective, appears to be separated and independent, but seen from a non-ordinary perspective, are still connected and form a whole. This means that, as it has been emphasized by Aerts in a number of papers~\cite{Aerts2, Aerts3, Aerts4, Aerts4b}, \emph{non-locality} of quantum (or quantum-like) entities is first of all a manifestation of \emph{non-spatiality} (see also the arguments presented in~\cite{Massimiliano,Massimiliano2,Massimiliano3}).  

This point being rather subtle, let us try to explore it a little further. For this, we can observe that our three-dimensional Euclidean space is a very specific ``theatre'' that we humans have isolated from the rest of reality, through the cognitive filters that emerged from our experiences with macroscopical entities. 

These macroscopic entities, Aerts explains, can be characterized by what he calls the property of \emph{macroscopic wholeness}. More precisely, quoting from~\cite{Aerts2}: 

\textbf{Macroscopic wholeness}. For macroscopic entities we have the following property: if they form a whole (hence are not two separated parts), then they hang together through space. Which means they cannot be localized in different macroscopically separated regions $R_1$ and $R_2$ of space, without also being present in the region of space `between' these separated regions $R_1$ and $R_2$. 

In other terms, from our ordinary spatial perspective, a composite (decomposable) entity exists as such, i.e., as a whole, for as long as its composing parts remain connected together \emph{through space}. However, we may ask if there are other possibilities in reality for the composing parts of an entity to remain connected together, apart from ``through space.'' 

Considering our $k$-model, we can observe that, although $S_K$ is disassembled during the measurements process, nevertheless, in the end, it gets \emph{necessarily} reassembled. This means that the fragments of $S_K$ remain (invisibly) dynamically connected through the specific structure of the measuring machine and, therefore, although temporarily spatially separated, they are nevertheless ``hanging together'' in a more subtle way.

The present discussion, as evident as it might appear, touches at the heart of our understanding of physical reality, and more particularly of our understanding of the fundamental concepts of spatiality and non-spatiality, actuality and potentiality, soft and hard act of creations, and macroscopic wholeness. As we are going now to explain, thanks to the intuitions we have gained from our $k$-model, all these concepts are in fact intimately related. Let us start with the concept of potentiality. 

\textbf{Potentiality}. As we have explained in the Introduction, a property is potential if it is not actual, i.e., if one cannot predict with certainty, even in principle, the ``yes'' outcome of the associated test. Potentiality however, can either be deterministic or indeterministic. More precisely, we shall say that a property is \emph{deterministically potential} if the ``no'' outcome can be predicted with certainty. On the other hand, we shall say it is \emph{indeterministically potential} (or, which is equivalent, \emph{indeterministically actual}) if it is neither actual nor deterministically potential, which means that the ``yes'' and ``no'' answers have both a certain propensity to manifest, but none of them can be predicted with certainty, i.e., the associated probabilities are strictly different from $0$ and $1$. 

For instance, for a classical particle, if transmissibility is actual, then reflectivity is deterministically potential, and vice versa. On the other hand, for a quantum entity, apart from the high and low energy regimes, transmissibility and reflectivity are indeterministically potential. Finally, in the quantum-like intermediate situations, both deterministic and indeterministic potentiality can be present in the system. 

Let us now consider what is the most fundamental property for any entity: \emph{existence}. In the case of $S_K$, we can identify such a property with the one of \emph{macroscopic wholeness}: entity $S_K$ exists if macroscopic wholeness is actual, i.e., if it forms a cohesive whole through space, in the sense defined by Aerts above. 

Clearly, at the beginning of the scattering experiment, the property of macroscopic wholeness of $S_K$ is actual, and one can say that $S_K$ is in its \emph{actual mode of being}. However, as soon as it falls in the inner compartment, and breaks in several pieces, its macroscopic wholeness becomes deterministically potential. Accordingly, one can say that $S_K$ enters into a \emph{potential mode of being}. And, as soon as the machine completes the measurement, macroscopic wholeness is restored, and the mode of existence of $S_K$ becomes actual again.

\section{Process-actuality}
\label{weak}

It is worth emphasizing that when we speak of the actuality or potentiality of a property (be it deterministic or indeterministic), it is always relative to a given moment of time (typically, the moment at which the outcome of the test that defines the property becomes available, if one would chose to perform it). 

Let us also observe that, as soon as $S_K$ is destroyed, during the measuring process, it literally disappears from our ordinary spatial perspective. But, if this is true, in what sense can we nevertheless say that $S_K$ still exists, although not in the actual sense? In other terms, does the statement ``$S_K$ is potentially existing'' have some objective correspondence in our reality, or is it just a way of saying, a heuristic statement that we must take care not to reify?

As we said, during the experiment, entity $S_K$ is temporarily disassembled, i.e., destroyed. Following Coecke's distinction between soft and hard acts of creations (see the Introduction) we can say that the machine performs, during the measurement, a succession of hard acts of creation on $S_K$ (and on its fragments), and although most of these acts are destructive, as they destroy the macroscopic wholeness of $S_K$, if taken all together they constitute in fact a soft act of creation, as is clear from the fact that $S_K$ is recreated in the final stage of the experiment, \emph{with certainty}. 

What is crucial to understand is that \emph{the machine acts in a deterministic way in the process of actualization of the potential existence of} $S_K$, in the sense that if we exclude anomalies (such as for example an experimenter who, by distraction, would hit the machine and make it fall to the ground while it is functioning), we can predict with certainty that at the end of the measurement $S_K$ will be reassembled. 

To put it another way, if nothing perturbs the functioning of the apparatus, the different fragments of $S_K$ will never lose their \emph{mutual coherence}, through the mediating structure of the machine, and thus remain (dynamically) connected \emph{through time} in such a way that their ``hanging together'' \emph{through space} will be guaranteed at the end of the process. 

These considerations lead us to define the new concept of \emph{process-actuality}. 

\textbf{Process-actuality}. A property is \textit{process-actual} (\emph{p-actual}), in a given moment, if it is actual in that moment or, if not, it will become actual in a subsequent moment, with certainty. In other terms, a property is p-actual at time $t$, if there exist a time $t'\geq t$, such that the property is actual at $t'$. \emph{N.B.}: actuality $\Rightarrow$ p-actuality.

Thanks to the notion of process-actuality, we can define the following properties: 

\textbf{Process-macroscopic wholeness}. An entity possesses the property of \emph{process-macroscopic wholeness} (\emph{p-macroscopic wholeness}), in a given moment, if the property of macroscopic wholeness is p-actual in that moment. \emph{N.B.}: macroscopic wholeness $\Rightarrow$ p-macroscopic wholeness.

\textbf{Process-existence}. An entity exists in the process sense (\emph{p-existence}), in a given moment, if its existence is p-actual in that moment. 
(This means, in particular, that some of the entity's intrinsic defining properties are p-actual in that moment). \emph{N.B.}: existence $\Rightarrow$ p-existence.

\section{Non-spatiality}
\label{non-spatiality}

In this section we exploit the process-actuality criterion to provide a clear definition and characterization of the important notion of non-spatiality. For this, we start by observing that \emph{existence} and \emph{spatiality} are intimately related concepts. Indeed, to exist is to exist in a given space, i.e., in the space to which belong the measuring apparatus that are used to test the properties and attributes of the entity under consideration.

Entities with different attributes can belong to a same space, and interact together in some way (by ``belonging to'' a space we don't only mean ``to be present in'' a space, but, more generally, ``to be detectable in'' a space; see in this regard the discussion in~\cite{Aerts4}, Sec. 2). On the other hand, within a same space, one can also identify subspaces, i.e., substructures that are characterized by the specific attributes of the entities that, by definition, belong to them. 

Considering our \emph{physical space}, we can certainly highlight in it an important subspace, that we can simply call the \emph{ordinary physical space}: 

\textbf{Ordinary physical space}. The \emph{ordinary physical space} ($S_{\texttt{or}}$) is that part of our physical space ($S_{\texttt{ph}}$) that contains entities for which the property of macroscopic wholeness is actual. \emph{N.B.}: $S_{\texttt{or}}\subset S_{\texttt{ph}}$.

Is $S_{\texttt{or}}$ isomorphic to the three-dimensional Euclidean space? That's possible, but not certain, as we cannot a priori exclude the existence in our physical space of, say, four-dimensional macroscopically whole entities (think about Abbott's metaphor of Flatland). Also, macroscopic wholeness may not be a sufficient condition to characterize $S_{\texttt{or}}$ as our 3-$d$ space, and other attributes may be needed for this. However, not to complicate the discussion, we shall assume in the following that $S_{\texttt{or}}$, as defined above, is indeed isomorphic to the three-dimensional Euclidean space. 

Let us now define what we shall call, for lack of a better term, the \emph{extraordinary physical space}:  

\textbf{Extraordinary physical space}. The \emph{extraordinary physical space} ($S_{\texttt{ex}}$) is that part of our \emph{physical space} ($S_{\texttt{ph}}$) that contains entities for which the property of macroscopic wholeness is p-actual. \emph{N.B.}: $S_{\texttt{or}}\subset S_{\texttt{ex}} \subset S_{\texttt{ph}}$.

\begin{figure}[!ht]
\centering
\includegraphics[scale =.6]{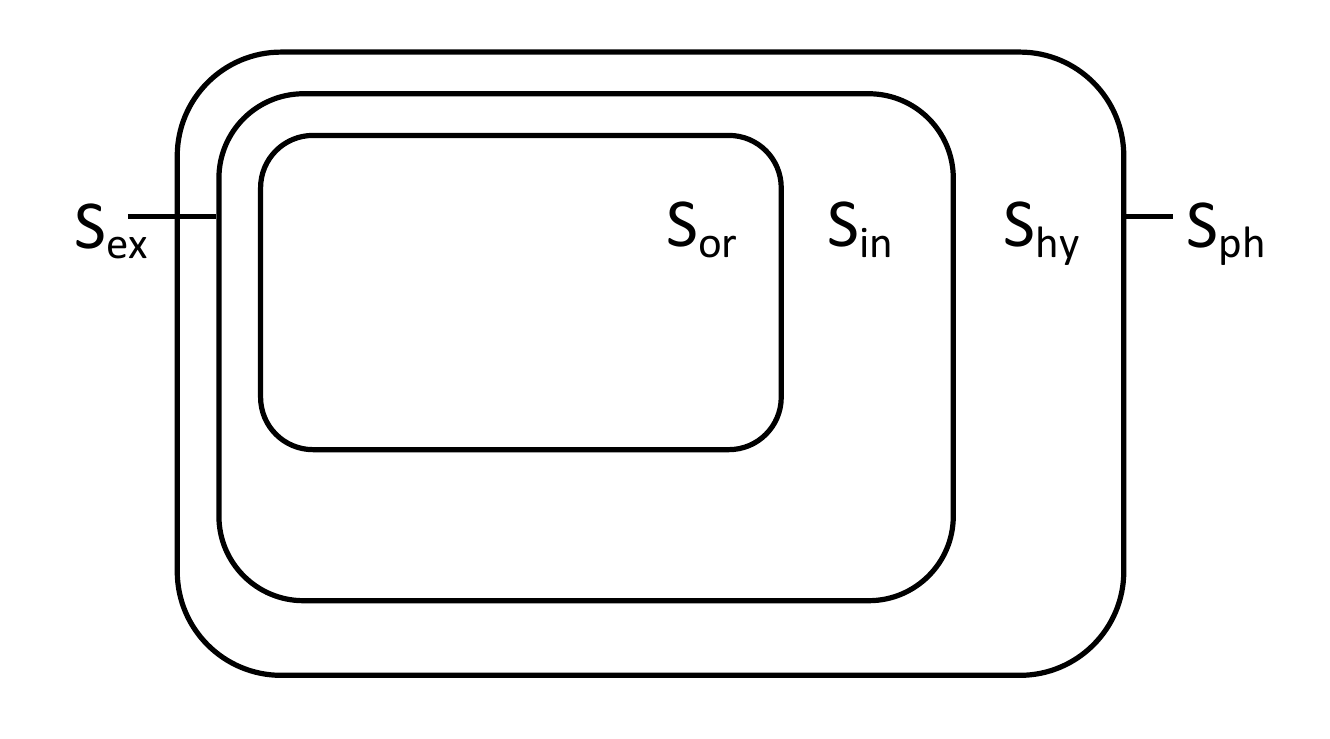}
\caption{A set-theoretical representation of the subspaces $S_{\texttt{or}}$, $S_{\texttt{in}}$, $S_{\texttt{ex}}$ and $S_{\texttt{hy}}$, of our physical space $S_{\texttt{ph}}$.
\label{spaces}}
\end{figure}

With the above definitions, we can also define the following two spaces (see Fig.~\ref{spaces}): 

\textbf{Intermediate physical space}. The \emph{intermediate physical space} ($S_{\texttt{in}}$) is that part of $S_{\texttt{ex}}$ that contains entities which are not in $S_{\texttt{or}}$, i.e., for which macroscopic wholeness is deterministically potential. In other terms, in a set-theoretical sense: $S_{\texttt{in}}= S_{\texttt{ex}}\backslash S_{\texttt{or}}$. 

\textbf{Hyperordinary physical space}. The \emph{hyperordinary physical space} ($S_{\texttt{hy}}$) is that part of $S_{\texttt{ph}}$ that contains entities which are not in $S_{\texttt{ex}}$, i.e., such that p-macroscopic wholeness is deterministically potential. In other terms, in a set-theoretical sense: $S_{\texttt{hy}}= S_{\texttt{ph}}\backslash S_{\texttt{ex}}$. \emph{N.B.}: the exact characterization of entities belonging to $S_{\texttt{hy}}$, if any, is unknown.

We are now in a position to propose a precise definition of non-spatiality, as this notion is conventionally used (also by the present author~\cite{Massimiliano,Massimiliano2,Massimiliano3}) in connection with quantum and quantum-like entities.

\textbf{Non-spatiality}. A non-spatial physical entity is, by definition, an entity that belongs to the intermediate physical space $S_{\texttt{in}}$. This means that non-spatiality is not a condition of absence of spatiality, but a condition of intermediate spatiality, such that ordinary spatiality and hyperodinary spatiality are absent. 

Let us illustrate the content of the above definitions, using the guiding example of entity $S_K$. At the beginning of the experiment, which as a whole is a soft act of creation, $S_K$ exists in $S_{\texttt{or}}$, as a macroscopically whole entity. Then, in the course of the experiment, it ceases to manifest in $S_{\texttt{or}}$, but is not for this totally annihilated, as it continues to exist in $S_{\texttt{in}}$, as a p-existing, p-macroscopic whole entity (a condition improperly called of non-spatiality). Then, at the end of the experiment, it manifests again in $S_{\texttt{or}}$, by acquiring once more the property of macroscopic wholeness.

Considering microscopic quantum entities, like for instance an electron, we are now equipped with some interesting conceptual tools that allow us to describe what might possibly happen between the preparation of the entity, at the beginning of a typical quantum measurement, and the ``click in the counter,'' at the end of it. 

If we assume that the conceptual framework we have so far explored with the help of our model is pertinent, we can think of a microscopic entity, like an electron, as a sort of composite entity. In some instances, when it is fully assembled in $S_{\texttt{or}}$, in a state of macroscopic wholeness, we are able to ``see'' it, with the ``eyes'' of our macroscopic instruments (which also belong to $S_{\texttt{or}}$). 

On the other hand, in some other instances, the electron-entity may disappear from our ``sight'', by losing its macroscopic wholeness, i.e., its wholeness through $S_{\texttt{or}}$, which then becomes a process-like, dynamical form of wholeness. In these moments, the electron only p-exists, but nevertheless still exists, within $S_{\texttt{in}}$, in a condition where its macroscopic wholeness is deterministically potential. 

Nevertheless, since the electron has not been destroyed, having been acted upon not by a hard act of creation, but by a soft act of creation (more precisely, by a succession of hard acts of creation whose overall effect results in a soft act of creation), it will finally demanifest from that ``non-spatial'' realm, to manifest again in our (three-dimensional) ordinary space, by restoring its macroscopic wholeness. 

Quoting Aerts from~\cite{Aerts4}: ``Reality is not contained within space. Space is a momentaneous crystallization of a theatre for reality where the motions and interactions of the macroscopic material and energetic entities take place. But other entities - like quantum entities for example - `take place' outside space, or - and this would be another way of saying the same thing - within a space that is not the three dimensional Euclidean space.''

According to our analysis, the space mentioned by Aerts in the above excerpt is our ordinary physical space ($S_{\texttt{or}}$), whereas the other space he mentions, that is not the three dimensional Euclidean space, is the intermediate space $S_{\texttt{in}}$, which is included in the larger physical space $S_{\texttt{ph}}$, and should certainly be considered as a part of our physical reality not less objective than $S_{\texttt{or}}$. 

But then, if this is so, why can't we see, with our eyes, this intermediate theatre of reality? A possible answer is because, in our construction of reality (and knowledge about reality) through the instrument of our highly noun-oriented language (particularly in Western countries), we have ended up developping a much more ``structure-oriented'' than ``process-oriented'' view. In other terms, we have developed more the tendency to observe reality as a collection of snapshots, rather than as a collection of continuous movies, each one endowed with its indeterministic aspects (related to our present and future acts of creations) and deterministic aspects (related to the effects of our past creations). 

Each one of these snapshots, or moments, creates the illusion of a static three-dimensional theatre, filled with ordinary objects, all characterized by the property of macroscopic wholeness. In other terms, by only creating our reality on ``instants,'' we generate the illusion of a ``snapshot-space,'' which we believe then to constitute a unique all inclusive theatre.

However, as we expand the consciential crack through which we look to the world (for instance by becoming aware of the AUUA, the Additional Unconsciously Used Assumption~\cite{Aerts2} that are present in our language and cognitive processes), we may realize that our rough cognitive filters are in fact screening us from the more dynamical (process-oriented) vision of the innumerable non-spatial entities, which are also objectively participating to our reality, although in a different mode of being. 

Let us point out that such a perceptual expansion is not about simply replacing our naive three-dimensional spatial theatre with an equally naive four-dimensional spacetime theatre, in which real change wouldn't at all be possible. However, discussing the very subtle aspects of the geometric and process views inherent in our construction of reality would go too far beyond the scope of the present work, and we refer the interested reader to Aerts' important contributions~\cite{Aerts4, Aerts12, Aerts13}.

\section{Final considerations}
\label{conclusion}

In this paper we have proposed a new quantum machine model, which is able to modelize simple classical, quantum and quantum-like one-dimensional scattering processes. Although our model is structurally much simpler than Aerts' $\epsilon$-model, it has the advantage of providing what we think is a suggestive metaphor for quantum entities. 

This because it allows to visualize what could happen when a quantum entity apparently disappears from our ordinary ``view'' and, in the period of time before it is detected again, becomes a genuine ``non-spatial'' entity, i.e., an entity which is not any more present in our ordinary physical space, but in an intermediate space, characterized by a teleological form of macroscopic wholeness.

Of course, there are many substantial differences between entity $S_K$ and a microscopic quantum entity, like an electron. Apart those we have already mentioned in the previous sections, there is the fact that the position of an electron is, generally speaking, an \emph{ephemeral property}~\cite{Massimiliano}, which can only remain actual for a moment. On the contrary, $S_K$ has clearly the ability to remain stably present in our ordinary space, for an arbitrary amount of time. 

According to the view expressed in this work, the ephemeral character of the position of an electron (or of any other microscopic entity) would be the consequence of the ephemeral character of its macroscopic wholeness. This, by the way, is also the reason why macroscopic wholeness is so called, and is not called, for instance, microscopic wholeness! 

An electron, contrary to $S_K$, expresses the preference to remain in a state of p-macroscopic wholeness (or possibly in a state of existence still different, corresponding to the hypothetical space $S_{\texttt{hy}}$) and it is only when ``forced'' by the action of a suitable measuring apparatus that it can acquire, for an instant, the property of macroscopic wholeness, being consequently detected in a given position of our ordinary space. In that respect, the quantum phenomenon known as the ``spreading of the wave packet,'' could very well be understood as a manifestation of this propensity of microscopic entities toward a more process-like form of wholeness and existence. 

Another interesting aspect revealed by our model is that p-existence would be strongly dependent on the action of the apparatus upon the entity's composing fragments. In other terms, contrary to existence in the ordinary sense, p-existence would be highly contextual, as without a specific measuring apparatus, able to coherently guide the evolution of the composing parts of a microscopic entity, the phenomenon of superposition and non-locality wouldn't probably be possible. 

In that sense, the very existence of microscopic entities, which most of their time are at best process-existing, would be much more contextual than the ordinary existence of macroscopic entities, which would express a more stable and context-independent condition of existence. 

To put it another way, not only the behavior of a quantum entity, like an electron, would depend on the nature of the questions we address to it, by means of our experiments, but also its possibility of p-existing would depend on the very presence of those processes that are embodied by the (coherence-preserving) experimental apparatus.

Of course, the nature of the influence exerted by a measuring apparatus on the components of a microscopic quantum entity is quite different from the interaction that is responsible of the final detection of the entity, for instance in the form of a little spot on a screen or a click in a counter. 

A question then arises: if, in a sense, it is possible to understand a microscopic quantum entity, like an electron, as a sort of compound entity, and if it is true that the components of such a compound entity will generally spread out while it interacts with the context made manifest by a measuring apparatus, how comes that we never directly detect the presence of these components? 

In other terms, if the $\delta$-quantum machine (or $k$-model) metaphor is not totally unfounded, why do we only observe the traces of the entire electron entity and never of its composing parts?

As we also discussed in the previous section, one can easily understand why we fail to detect the electron-entity, as it evolves inside the experimental apparatus: being in a ``spread out'' state, it only p-exists from the view point of our ordinary physical space. This is what our quantum machine model suggests: when $S_K$ is disassembled, although its composing fragments remain correlated through time, thanks to the mediation of the experimental apparatus, it disappears from our object-oriented $S_{\texttt{or}}$-perspective, characterized by macroscopic wholeness.
 
However, in our quantum machine model, we can nevertheless directly observe, if we open the machine's box, the tiny spheres forming $S_K$, also when they are spatially separated. The reason for this is that the spheres have a double level of existence: they exist as the correlated components of a p-existing compound entity, but they also exist as individual entities. In other terms, each one of the composing elements of $S_K$ owns, in turn, the attribute of macroscopic wholeness, which is the reason why they can also be individually detected in our three-dimensional space. 

On the other hand, the situation of a microscopic entity, like an electron, would be different. Indeed, its elementariness would prevent it from being decomposed in sub-entities that would in turn still possess the attribute of macroscopic wholeness. This suggests to define \emph{elementariness} not as the property of an entity of not being made up of other entities, as one usually does, but as the property of an entity of not being decomposable into sub-entities that would also possess, in turn, the property of macroscopic wholeness. More precisely, in the speculative logic of the present work, we propose the following definition of elementariness:

\textbf{Elementariness}. An entity is \emph{elementary} if it can actualize, at least ephemerally, the property of macroscopic wholeness, whereas its composing parts cannot. In other terms, an elementary entity is an entity that can be detected in $S_{\texttt{or}}$, whereas its components are  confined in $S_{\texttt{ph}}\backslash S_{\texttt{or}}$.

Considering the above proposed definition, elementariness would not be about being or not being decomposable (as every entity can be assumed to always be decomposable, until proven to the contrary), but about the impossibility for the composing fragments to belong, even ephemerally, to $S_{\texttt{or}}$.

The difficulty we have in visualizing the above concept of elementariness, resides in the fact that we have the tendency to think about the composing parts of an entity in corpuscular terms. And this is because most of our visualization tools are inherited from our three-dimensional experience of the macro-world, i.e., from our experience with so-called ordinary \emph{objects}, which are entities possessing, in a stable way, the property of macroscopic wholeness. 

Clearly, this is where the metaphor of our quantum machine model ceases to be helpful in guiding our intuition. An electron-entity is not an ordinary object, and although we can imagine it as being decomposable, its composing parts are not in themselves elementary, but \emph{pre-elementary}, as they strictly belong to a space which is beyond our ordinary level of experience.   

Nonetheless, can we devise a way to detect these non-ordinary composing parts, these hypothetical sub-elementary ``partons,'' of an elementary entity? A natural way to proceed would be to design experimental apparatus whose functioning would not be limited to $S_{\texttt{or}}$. Still, if we reflect attentively, we may realize that these non-ordinary machineries already exist, and could be nothing more than the sophisticated devices already present in our modern physics laboratories. 

These advanced instruments~\footnote{As a paradigmatic example, consider a perfect silicon crystal, as used in neutron interferometric experiments, where single neutrons can simultaneously manifest in separated regions of the crystal, then recombine in the detectors, giving rise to typical interference patterns ~\cite{Rauch}.} have indeed been carefully designed to reveal the quantum properties of physical entities, i.e., to highlight the hidden and subtle connections between the sub-elementary composing parts of microscopic quantum entities, which are responsible of the observed non-local and superposition interference effects, typical of our quantum level of reality.

Of course, from our classical, ordinary viewpoint, it may not be easy to accept such evidence, and we could be tempted to believe that in quantum experiments we can never say what actually goes into them, and can only comment about their outputs. Outputs, no doubts, are easier to comment, as they belong to $S_{\texttt{or}}$. However, if we observe the functioning of a typical quantum experiment with a more process-oriented perspective, we may conclude that, in fact, they do reveal much more than we are usually led to believe. 

To give an example, the so-called ``wave-particle duality,'' as observed in a double-slit experiment, could be seen as an expression of the hidden connections that are present among the different components of an elementary entity, when in its process mode of being. Waves indeed, can be understood as phenomena resulting from the coherent collective movement of a great number of correlated entities, forming the medium through which the wave-perturbation is said to propagate. 

But of course, the great difference between a classical wave and a quantum wave-like phenomenon lies in the fact that the former is a perturbation manifesting in our ordinary three-dimensional physical space, whereas the latter doesn't. 

This bring us back to our mentioned cognitive blindness, in seeing what a quantum measurement really reveals us; a blindness related to our bad habit of thinking to the entities populating our reality only in terms of corpuscles, or classical fields and waves. These images, as useful as they may be in the description of macroscopic entities, are nevertheless totally misleading if we use them in the description of microscopic ones (be them elementary, like an electron, or non-elementary, like an atom or a molecule). 

But then, what would be a better notion to properly think about quantum or quantum-like entities? A fascinating answer comes from Aerts' recent proposal to interpret quantum entities as... \emph{conceptual entities}! Indeed, according to Aerts, quantum entities would~\cite{Aerts15} ``[...] interact with ordinary matter, nuclei, atoms, molecules, macroscopic material entities, measuring apparatus,..., in a similar way to how human concepts interact with memory structures, human minds or artificial memories.'' 

It's not our intention to go here into the details of this subtle explanatory framework and the interesting path that led its author to develop it, and refer the interested reader to Aerts' thought provoking articles~\cite{Aerts15, Aerts16}. Let us however use this interpretation to highlight one of the ideas we have put forward in this paper, inspired by our machine-model: the compoundness of an elementary particle.

For this, let us consider, as an example, the conceptual entity called ``apple.'' Clearly, such a conceptual entity can manifest in our ordinary physical space, each time that a physical apple comes into being. If, for simplicity, we assume that apples ripen only in a very specific and short period of the year, and that soon after they are all eaten, then, similarly to an electron, we can say that an apple conceptual entity will live most of its time outside of our ordinary space, and just briefly enter it, when the great spring experiment is performed. 

Now, considering an apple-object, which is the objectification of an apple conceptual entity (in the same way as the spot we observe on a detection screen can be considered as the objectification of an electron conceptual entity), we can easily think of it as a compound entity, made of other objects, like for instance its peel, pulp, seeds, stem, and so on. Clearly, all these connected parts individually belong to $S_{\texttt{or}}$, as the whole apple-object does. 

But what about the apple conceptual entity, can we also understand it as a compound entity? Consider for instance the concepts ``typical'' and ``fruit''. These two concepts, contrary to the apple-concept, cannot be also understood as objects. Indeed, there are no objects in $S_{\texttt{or}}$ corresponding to ``typical'' and ``fruit'' (on the shelves of a grocery store one finds apples, pears, oranges, etc., but not the fruits called fruit!). 

Considering however the connection (through meaning) of the ``typical'' and ``fruit'' concepts, we obtain the composed concept ``typical fruit,'' which, for almost every person, is nothing but an apple. In other terms, we have an example of a conceptual entity, an apple, which, from time to time, can manifest as an object in $S_{\texttt{or}}$, and which can be understood as the combination two other conceptual entities, ``typical'' and ``fruit,'' that instead cannot manifest inside of $S_{\texttt{or}}$.

This perfectly illustrates our above notion of elementariness. Similarly to the apple conceptual entity, the electron conceptual entity can be understood as the composition (combination) of other conceptual entities, which, however, cannot manifest, not even ephemerally, in $S_{\texttt{or}}$.

Of course, the ``apple'' example is not a perfect example of an elementary conceptual entity, as it is also possible to decompose it in parts which are objects. An electron on the other hand, would be truly elementary because the only possible decompositions would be of the ``typical-fruit'' kind, and not of the ``peel-pulp'' kind. (In that sense, the apple example better describes an atom, or a molecule, than an elementary entity like an electron).

Much more should certainly be said about the conceptual status of quantum entities, to truly appreciate the explicative power of this interpretation, recently developed by Aerts. Also, much more should be said about the many fundamental notions we have just touched upon in this article, many of which are quite speculative and certainly deserve a larger space of analysis. We hope this larger space will be available in future works.

Before concluding, a word of warning is due. As we have seen, our machine model is quite evocative and seems to suggest that, in some way, quantum entities could be understood as some sort of composite entities, a view that, as we have briefly explained, is not incompatible with Aerts' conceptual interpretation of quantum mechanics.

Nevertheless, we would like to point out (especially for the hasty reader that would have been left with a wrong impression) that the $\delta$-quantum machine model is not meant to suggest that quantum entities should be considered, in a literal sense, as composite entities made of minuscule undetectable classical (Bohomiam-like) particles, able to spread all over the space. This no more than Aerts' spin-quantum machine model is meant to suggest that there is a real breakable elastic band hidden somewhere in a Stern-Gerlach magnet!

The truly interesting aspect about these models is not their ability to realistically describe physical entities as such, but to capture, by means of powerful structural analogies, the possible logic that would be at the basis of their interaction with the different experimental contexts, particularly for what concerns the emergence of quantum probabilities. 

On that purpose, it is important to observe that Aerts' spin-quantum machine is able to capture the essence of the quantum probability structure without any need to assume that the entity under investigation has a composite structure. Also, there are certainly ways to adapt the spin-quantum machine model to also characterize one-dimensional quantum scattering processes, as is clear from the fact that there is an isomorphism between a spinor and the two-component column vector representing an incoming/outgoing one-dimensional scattering state (for a given energy). 

If we are saying all this, is to bring the reader to consider that the composite nature of entity $S_K$, in our $\delta$-quantum machine model, is not necessarily a fundamental logical ingredient in the description of a quantum entity. Of course, \emph{compoundness} seems to be important at some level of the description, as for instance also in the spin-quantum machine model the elastic is a composite entity, that can be disassembled (broken) in different ways. But we don't know if this ``compoundness-breakability'' property is a \emph{sine qua non} ingredient in the quantum description of reality and, if so, at what level it should be applied (at the level of the entity, of the measuring apparatus, of both, etc.).

In other terms, our words of caution are to point out that the really profound aspects revealed by the different machine models is probably not in what distinguishes them, in terms of details, but in what they have in common, at a structural level, like for instance a built-in mechanism that selects (in a highly contextual way) a deterministic ``hidden'' measurement, and the possibility of creating new properties during its execution. 

It is now time, to conclude, and leave the last word to Aristotle, whose idea of causality, in his theory of movement (and, more generally, of transformation) expresses, in a way, the idea of process-existence that we have put forward in the present article. Quoting from~\cite{Smets}: 

``[...] \emph{everything that comes to be moves towards a principle, i.e. an end. For that for the sake of which a thing is, is its principle, and the becoming is for the sake of the end; and the actuality is the end, and it is for the sake of this that the potentiality is acquired.}''

\begin{acknowledgements}
I'd like to express thanks to two anonymous referees, for the attentive reading of the manuscript and their helpful comments, which have contributed in significantly improving its presentation. I'm also grateful to Diederik Aerts, for his support and interest about the subject of the present article. 
\end{acknowledgements}

\end{document}